\documentclass[11pt]{article}
\usepackage{amsmath,amssymb,amsthm}
\usepackage{graphicx}
\usepackage{longtable,booktabs,array}
\usepackage{subcaption}
\usepackage[round]{natbib}
\usepackage{placeins,needspace,xcolor}
\usepackage[hidelinks]{hyperref}
\newtheorem{theorem}{Theorem}
\newtheorem{lemma}[theorem]{Lemma}
\newtheorem{corollary}[theorem]{Corollary}
\newtheorem{proposition}[theorem]{Proposition}
\newtheorem{condition}[theorem]{Condition}
\theoremstyle{definition}
\newtheorem{definition}[theorem]{Definition}

\newtheorem{remark}[theorem]{Remark}

\newcommand{\R}{\mathbb{R}}
\newcommand{\cB}{\mathcal{B}}
\newcommand{\cE}{\mathcal{E}}
\newcommand{\cF}{\mathcal{F}}
\newcommand{\cG}{\mathcal{G}}
\newcommand{\cK}{\mathcal{K}}
\newcommand{\cM}{\mathcal{M}}

\newcommand{\argmin}{\operatorname*{argmin}}
\newcommand{\diag}{\operatorname{diag}}
\newcommand{\diam}{\operatorname{diam}}

\input{supplement_numbers}

\begin{document}

\title{Tree-aggregated compositional regression under measurement error}

\author{
Zhenghan Li\textsuperscript{1} and
Tianying Wang\textsuperscript{2}\thanks{Corresponding author. Email: Tianying.Wang@colostate.edu}\\[0.5em]
\textsuperscript{1}School of Statistics, University of Minnesota\\
\textsuperscript{2}Department of Statistics, Colorado State University
}

\date{}

\maketitle
\thispagestyle{empty}
\baselineskip=20pt

\begin{abstract}
Compositional data arise in many applications, including microbiome studies, where taxonomic or phylogenetic trees describe relationships among taxa. These trees allow taxa with a common regression coefficient to be modeled as a group. An association may be specific to an individual genus or shared across a broader taxonomic group. Measurement error can make an association shared across a group appear specific to smaller groups, even when coefficient signs are preserved. We develop Tree-Aggregated Regression with Correction for Observation Error, or TARCO, to estimate latent coefficients and identify groups along the tree. We establish finite-sample conditions for exact recovery of the true subtree groups and derive bounds on prediction and coefficient estimation errors. The guarantees extend to settings where the measurement error covariance is estimated from auxiliary replicates. Simulations show improvements in prediction and coefficient estimation over tree aggregation without measurement error correction, and illustrate how the choice of tree penalty affects group recovery. In a gut microbiome analysis, TARCO identifies shared associations with body mass index and shows how correction changes their relative strengths across taxonomic groups.
\end{abstract}


\pagestyle{plain}
\begingroup
\allowdisplaybreaks

\section{Introduction}

Compositional regression has long been used to study associations between an outcome and predictors measured as proportions \citep{aitchison1984log, lin2014variable}. Hierarchies among predictors have also been used in statistical regression to guide variable selection and aggregate related predictors \citep{zhao2009composite, yan2021rare}. For compositional predictors, a prespecified tree places the components at its leaves and defines nested groups at its internal nodes. Taxonomic and phylogenetic trees provide such hierarchies in microbiome studies, where an association may be specific to one genus or shared across a larger taxonomic group. When the components in a subtree share a coefficient, a single regression term can represent their shared effect even if many leaf coefficients are nonzero. Tree-guided regression uses this property to learn the level of aggregation from the data \citep{bien2021tree, li2023s}.

Measurement error is common in microbiome studies, where relative abundances are estimated from sequencing reads. For example, differences in how efficiently taxa are measured can cause the observed proportions to differ from the underlying composition \citep{brooks2015truth, mclaren2019consistent}. Random errors in the observed log ratios can also bias coefficient estimates \citep{shi2022high}. Gram matrix correction accounts for measurement error in high-dimensional regression \citep{loh2012high}. CoCoLasso introduced a positive semidefinite projection to make the corrected quadratic loss convex, and subsequent methods use this construction with other penalties \citep{datta2017cocolasso,zheng2018balanced}. For compositional predictors, Error-in-Composition Lasso (Eric Lasso) provides guarantees for coefficient estimation and sign recovery in log-contrast models \citep{tan2024high}. Tree-guided regression also requires identifying which components share a common latent coefficient. We show that taxa sharing a coefficient in the latent regression can have different coefficients in the population regression on the observed composition. An association shared across a subtree may then appear specific to smaller subtrees, even when all coefficient signs are preserved (see the three-component example in Section~\suppref{sec:toy_example_appendix} of the Supplement). We ask when the maximal subtrees with a shared latent coefficient can be recovered from error-prone compositions.

We develop Tree-Aggregated Regression with Correction for Observation Error (TARCO) to estimate shared regression effects along a tree from error-prone compositions. TARCO corrects the quadratic loss for measurement error and estimates the latent log-contrast coefficients using a tree penalty that favors shared effects within subtrees. Simulations show improvements in prediction and coefficient estimation over tree aggregation without measurement error correction, and illustrate how penalty choice affects group recovery. In the gut microbiome analysis, correcting measurement error changes the relative strengths of associations with body mass index (BMI) across the taxonomic tree.

We show when the latent subtree partition can be recovered from error-prone compositions. Our finite-sample guarantee holds when the separation between groups is sufficiently large relative to the largest coefficient estimation error. The fitted coefficients within a true group may differ, yet thresholding the coefficient range within each subtree at a suitable level can recover the latent subtree partition exactly. Our theory therefore addresses the level at which an association is shared, in addition to individual coefficient accuracy. We also bound prediction error and extend the guarantees to settings where the error covariance is estimated from auxiliary technical replicates. The rest of the paper is organized as follows. Sections~\ref{sec:method} and~\ref{sec:theoretical_results} present the method and theory. Sections~\ref{sec:simulation_studies} and~\ref{sec:applications} report the simulations and application, followed by the discussion.

\Needspace{6\baselineskip}
\section{Method}
\label{sec:method}

We estimate the log-contrast coefficient vector $\beta^*$ and use the tree to represent effects shared by related taxa.
TARCO corrects the observed Gram matrix and applies a positive semidefinite projection weighted by leaf counts. The projection preserves the quadratic loss across equivalent tree representations.

\subsection[Model setup and tree parameterization]{Model setup and tree parameterization}
\label{sec:model_setup}

Denote by $X=(X_{ij})$ an $n\times p$ latent compositional data matrix with $X_{ij}>0$ and $\sum_{j=1}^{p}X_{ij}=1$, and by
$y=(y_1,\ldots,y_n)^\top$ an outcome vector. We use the additive log-ratio
(ALR) transformation with the $p$th component as reference
\citep{aitchison1982statistical} and write
\[
Z_{ij}
=
\log(X_{ij}/X_{ip})
=
\log(X_{ij})-\log(X_{ip}),
\qquad j=1,\ldots,p-1.
\]
For notational convenience, we include the reference coordinate $Z_{ip}=0$ and write $Z\in\R^{n\times p}$. For the theoretical analysis, we center the response and log-ratio covariates at their population means, so the model below has no intercept. The numerical analyses include an unpenalized intercept through sample centering, as described in Section~\ref{sec:penalty_computation}. The corresponding log-contrast model is \citep{aitchison1984log,lin2014variable,shi2016regression}
\begin{equation}
y_i
=
\sum_{j=1}^{p-1}\beta_j^*Z_{ij}+e_i,
\qquad i=1,\ldots,n.
\label{eq:logratio_linear}
\end{equation}
Here, $e_i$ denotes the regression error, and we write $e=(e_1,\ldots,e_n)^\top$. We use $Z_{\cdot j}$ for the $j$th column of $Z$.
The reference coefficient is determined from
the first $p-1$ coefficients through the constraint
$\mathbf 1_p^\top\beta^*=0$.
Write $\cB=\{b\in\R^p:\mathbf 1_p^\top b=0\}$ for the contrast space.

Let $\mathcal T$ be a prespecified tree with $T$ indexed non-root nodes and
leaves $L(\mathcal T)=\{1,\ldots,p\}$. For each node $k$, let $L_k$ be its
descendant leaf set and write $w_k=|L_k|$. Define the aggregation matrix
$A\in\{0,1\}^{p\times T}$ by $A_{jk}=1$ if node $k$ is an ancestor of leaf
$j$, and $A_{jk}=0$ otherwise. Each leaf is its own ancestor, so the leaf
columns of $A$ contain the $p\times p$ identity matrix. In particular,
$\operatorname{rank}(A)=p$.

For a coefficient vector $\beta^*$ satisfying
$\mathbf 1_p^\top\beta^*=0$, define its feasible tree representations by
\[
\cF_{\mathcal T}(\beta^*)
=
\left\{
\gamma\in\R^T:
A\gamma=\beta^*,
\ (A^\top\mathbf 1_p)^\top\gamma=0
\right\}.
\]
For every $\gamma\in\cF_{\mathcal T}(\beta^*)$, the model has the
equivalent representation
\[
y
=
Z\beta^*+e
=
ZA\gamma+e
=
\sum_{k=1}^T\left(\sum_{j\in L_k}Z_{\cdot j}\right)\gamma_k+e.
\]
The identifiable coefficient target is $\beta^*$. The vector $\gamma$ is an
auxiliary tree representation and is generally not unique.

In the error-free setting, let $\mathcal P(\gamma)$ denote a tree penalty and $\lambda>0$ its tuning parameter. The penalty choices are described in Section~\ref{sec:penalty_computation}. We consider the following penalized log-contrast objective:
\[
\frac{1}{2n}\|y-Z\beta\|_2^2+\lambda\mathcal P(\gamma),
\qquad
\mathbf 1_p^\top\beta=0,
\quad
\beta=A\gamma,
\]
or equivalently, after dropping the constant $y^\top y/(2n)$,
\[
\frac12\gamma^\top\Psi\gamma-\psi^\top\gamma
+\lambda\mathcal P(\gamma),
\qquad
(A^\top\mathbf 1_p)^\top\gamma=0,
\]
where
$\Psi=n^{-1}A^\top Z^\top ZA$ and
$\psi=n^{-1}A^\top Z^\top y$.

\subsection[Measurement error model]{Measurement error model}
\label{sec:error_model}

Let $\widetilde X=(\widetilde X_{ij})$ denote the observed composition.
Suppose the pre-compositional variables satisfy
$\widetilde R_{ij}=R_{ij}v_{ij}$,
where $R_{ij}$ is the latent pre-compositional value and $v_{ij}>0$ is a
multiplicative error (\citealp{firth2023analysiscompositionoriginalscale}; \citeauthor{tan2024high}, \citeyear{tan2024high}). With
$R_{i+}=\sum_{j=1}^pR_{ij}$ and
$\widetilde R_{i+}=\sum_{j=1}^p\widetilde R_{ij}$, define
$X_{ij}=R_{ij}/R_{i+}$ and
$\widetilde X_{ij}=\widetilde R_{ij}/\widetilde R_{i+}$. Thus
$\widetilde X_{ij}=X_{ij}v_{ij}R_{i+}/\widetilde R_{i+}$, and both the
latent and observed compositions sum to one. Applying the additive log-ratio
transformation gives
\[
\widetilde Z_{ij}=Z_{ij}+U_{ij},
\qquad
U_{ij}=\log(v_{ij}/v_{ip}),
\qquad j=1,\ldots,p-1.
\]
We extend the variables by setting
$Z_{ip}=U_{ip}=\widetilde Z_{ip}=0$. Thus
$Z,U,\widetilde Z\in\R^{n\times p}$ and
$\widetilde Z=Z+U$.

The stochastic analysis assumes that the rows
$\{(Z_i,U_i,e_i)\}_{i=1}^n$ are independent, that
$\mathbb E(U_i)=0$, that $U_i$ is independent of $(Z_i,e_i)$, and that
$\mathbb E(U_iU_i^\top)=\Sigma_U^{(p)}$. The last row and column of
$\Sigma_U^{(p)}$ are zero under the embedded reference convention. If
$\Sigma_U$ denotes the covariance of the first $p-1$ additive log-ratio
errors, then
$\Sigma_U^{(p)}=(\mathbf I_{p-1},\mathbf 0_{p-1})^\top
\Sigma_U(\mathbf I_{p-1},\mathbf 0_{p-1})$.
The scale invariance of the multiplicative construction follows directly
because multiplying all $v_{ij}$ within an observation by a common positive
constant leaves every ratio $v_{ij}/v_{ip}$ unchanged.

For tree-aggregated data, write
$Z_A=ZA$, $U_A=UA$, and $\widetilde Z_A=\widetilde ZA$.
The covariance of an aggregated error row is
$\Sigma_{U,A}=A^\top\Sigma_U^{(p)}A$.
When auxiliary replicate measurements are available, differences within each unit
can be used to form a preliminary estimate of $\Sigma_U^{(p)}$
\citep{carroll2006measurement}. Section~\ref{sec:replicate} defines the
estimator used here and gives its probability bounds.

\subsection{Beta-level tree groups and error amplification}
\label{sec:aggregation_process}

The tree plays two roles. It defines the coefficient groups used for
interpretation and determines how measurement errors at the leaves accumulate.
For $b\in\R^p$ and a tree node $k$, define
$\diam_b(L_k)=\max_{i,j\in L_k}|b_i-b_j|$.

\Needspace{5\baselineskip}
\begin{definition}[Beta-level tree groups]
\label{def:aggre_process}
Let $\mathcal V(\mathcal T)$ denote all nodes of the tree, including the
root, and define
\[
\cK^*
=
\left\{
k\in\mathcal V(\mathcal T):
\begin{array}{l}
\diam_{\beta^*}(L_k)=0,\\
\text{there is no strict ancestor }a\succ k\\
\text{ with }\diam_{\beta^*}(L_a)=0
\end{array}
\right\}.
\]
The beta-level tree groups are
$\cG^*=\{L_k:k\in\cK^*\}$.
\end{definition}

Any two descendant leaf sets are either disjoint or one contains the other.
Therefore, the maximal constant subtrees in $\cG^*$ partition the leaves.

\begin{figure}[htbp]
    \centering
    \includegraphics[height=4cm]{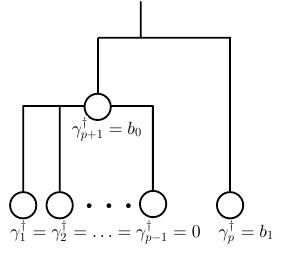}
    \caption{Tree aggregation with coefficients $b_0$ on the first $p-1$ leaves and $b_1=-(p-1)b_0$ on leaf $p$. Node labels give a representation $\gamma^\dagger$.}
    \label{fig:eg_AP}
\end{figure}

If the observed design is substituted directly for $Z$, the naive aggregated
quantities are
\[
\Psi_{\mathrm{naive}}
=
n^{-1}A^\top\widetilde Z^\top\widetilde ZA,
\qquad
\psi_{\mathrm{naive}}
=
n^{-1}A^\top\widetilde Z^\top y.
\]
The naive Gram matrix is biased because
\[
\mathbb E(n^{-1}\widetilde Z^\top\widetilde Z)
=
\mathbb E(n^{-1}Z^\top Z)+\Sigma_U^{(p)}.
\]
Tree aggregation further makes the stochastic scale depend on descendant-set
sizes. The following proposition records this entrywise effect.

For a scalar random variable $V$, write
$\|V\|_{\psi_2}=\inf\{d>0:\mathbb E\exp(V^2/d^2)\leq2\}$.

\begin{proposition}[Tree-dependent error amplification]
\label{Pr:heter_var}
There exist positive constants $c_2,c_3$ such that the following holds.
Assume that the pairs $(Z_i,U_i)$ are independent across $i$, that
$\mathbb E(U_i)=0$, that
$\mathbb E(U_iU_i^\top)=\Sigma_U^{(p)}$ for every $i$, and that $U_i$ is
independent of $Z_i$. Uniformly in $i$, suppose
\[
\max_{1\leq j\leq p}\|Z_{ij}\|_{\psi_2}\leq\nu_Z,
\qquad
\max_{1\leq j\leq p}\|U_{ij}\|_{\psi_2}\leq\nu_U.
\]
Let $\widetilde\sigma_{kl}$ be the $(k,l)$th entry of $\Sigma_{U,A}$.
For every node pair for which the following interval is nonempty, and for
every
\[
0<\epsilon\leq
\min\{c_2w_kw_l\nu_U\nu_Z,\widetilde\sigma_{kl}/2\},
\]
we have
\[
\begin{split}
&P\left(|\Psi_{\mathrm{naive},kl}-\Psi_{kl}|>\epsilon\right)\\
&\quad\geq
1-6\exp\left[
-\frac{c_3n\min\{(2\epsilon-\widetilde\sigma_{kl})^2,\epsilon^2\}}
{w_k^2w_l^2\max\{\nu_U^2\nu_Z^2,\nu_U^4\}}
\right].
\end{split}
\]
\end{proposition}

The deviation scale in Proposition~\ref{Pr:heter_var} depends on
both $\widetilde\sigma_{kl}$ and $w_kw_l$. This dependence motivates the
weighting by leaf counts used below. The proof is given in Section~\suppref{sec:error_amplification_proof}. Section~\suppref{sec:toy_example_appendix} gives calculations for the tree in Figure~\ref{fig:eg_AP} and a separate example showing how measurement error can break coefficient equality within a subtree.

\subsection{Corrected tree-aggregated estimator}
\label{sec:TARCO}

TARCO constructs a convex quadratic loss for tree-aggregated regression by correcting the Gram matrix and applying a positive semidefinite projection
\citep{datta2017cocolasso}. We formulate the projection in tree coordinates,
with weighting by leaf counts and a constraint that gives the same loss to equivalent tree representations. Define the corrected leaf-space Gram matrix
and score by
\[
\widehat Q
=
\frac1n\widetilde Z^\top\widetilde Z-\widehat\Sigma_U^{(p)},
\qquad
\widetilde q
=
\frac1n\widetilde Z^\top y,
\]
where $\widehat\Sigma_U^{(p)}$ is the covariance supplied to TARCO, either the
known covariance $\Sigma_U^{(p)}$ or the preliminary auxiliary-replicate
estimate defined in Section~\ref{sec:replicate}. The corresponding
tree-space quantities are $\widehat\Psi=A^\top\widehat QA$ and
$\widetilde\psi=A^\top\widetilde q$.
The score $\widetilde q$ requires no covariance subtraction because
$U_i$ is independent of $y_i$ and has mean zero.

Covariance subtraction may make $\widehat\Psi$ indefinite in finite samples.
Let the columns of $N\in\R^{T\times(T-p)}$ span $\ker(A)$, with the
constraint below understood as vacuous when $T=p$. Define
$W=\diag(w_1,\ldots,w_T)$ and use the kernel-preserving projection
\begin{equation}
\widetilde\Psi
\in
\argmin_{M\succeq0,\ MN=0}
\left\|W^{-1}(\widehat\Psi-M)W^{-1}\right\|_{\max},
\label{eq:kernel_projection}
\end{equation}
where $\|B\|_{\max}=\max_{k,l}|B_{kl}|$. The linear constraint ensures
that the corrected quadratic loss is the same for all tree representations
$\gamma$ satisfying $A\gamma=b$.

Because $A$ has row rank $p$, every symmetric matrix satisfying $MN=0$ has
a unique factorization $M=A^\top QA$. Positive semidefiniteness of $M$ is
equivalent to positive semidefiniteness of $Q$. The leaf-count weights also give
$\|W^{-1}A^\top HAW^{-1}\|_{\max}=\|H\|_{\max}$ for every symmetric $H$.
Thus \eqref{eq:kernel_projection} is equivalent to
\begin{equation}
\widetilde Q\in\argmin_{Q\succeq0}\|\widehat Q-Q\|_{\max},
\qquad \widetilde\Psi=A^\top\widetilde QA.
\label{eq:leaf_projection_main}
\end{equation}
This projection has a solution set independent of the tree
for a fixed leaf-space input and ALR reference. Existence of the projection and the equivalences above are established in
Section~\suppref{sec:projection}.

For fixed $\alpha\in\R$ and $\lambda>0$, consider the weighted tree penalty
\[
\mathcal P_\alpha(\gamma)
=
\|W^\alpha\gamma\|_1
=
\sum_{k=1}^Tw_k^\alpha|\gamma_k|.
\]
The corrected gamma-space estimator is
\begin{equation}
\widehat\gamma
\in
\argmin_{(A^\top\mathbf 1_p)^\top\gamma=0}
\left\{
\frac12\gamma^\top\widetilde\Psi\gamma
-\widetilde\psi^\top\gamma
+\lambda\|W^\alpha\gamma\|_1
\right\},
\qquad
\widehat\beta=A\widehat\gamma.
\label{eq:TARCO}
\end{equation}
The parameter $\alpha$ controls the relative cost of representations at
different tree levels. The theory below concerns this weighted $\ell_1$
choice.

Define the induced beta-space penalty
\[
\Omega_\alpha(b)
=
\min_{\gamma:A\gamma=b}\|W^\alpha\gamma\|_1,
\qquad b\in\R^p.
\]
The minimum is attained, $\Omega_\alpha$ is a norm, and its dual norm is
$\Omega_\alpha^*(z)=\|W^{-\alpha}A^\top z\|_\infty$.
Using $\beta=A\gamma$ and the induced penalty $\Omega_\alpha$, we can write
\eqref{eq:TARCO} equivalently as
\begin{equation}
\widehat\beta
\in
\argmin_{\mathbf 1_p^\top b=0}
\left\{
\frac12b^\top\widetilde Qb
-\widetilde q^\top b
+\lambda\Omega_\alpha(b)
\right\}.
\label{eq:beta_estimator}
\end{equation}
More precisely, the beta minimizers are exactly the images under $A$ of the
gamma minimizers, and every gamma minimizer is a penalty-minimizing
representation of its beta image. Section~\suppref{sec:induced_norm} establishes these norm properties and the exact correspondence between the two minimizer sets, even when the minimizers are not unique.

\subsection{Penalty choice and computation}
\label{sec:penalty_computation}

The primary penalty is the weighted $\ell_1$ penalty defined in
\eqref{eq:TARCO}. Setting $\alpha=0$ gives the standard $\ell_1$ penalty on $\gamma$.
Positive values of $\alpha$ assign larger penalties to nodes with more
descendant leaves, whereas negative values assign relatively larger penalties to
nodes with fewer descendant leaves. The induced norm $\Omega_\alpha$ makes clear that these
weights act on the identifiable coefficient through the least costly tree
representation.

For a constant coefficient $c$ on a subtree with $w$ leaves, the
representation using a single node costs $w^\alpha|c|$, compared with $w|c|$
for the representation supported on the leaves. For $\alpha\geq1$,
$\Omega_\alpha(b)=\|b\|_1$, giving a matched estimator without tree regularization when the
corrected loss and tuning procedure are held fixed. The proof is given in Proposition~\suppref{prop:penalty_boundary_supp}. This weighted tree penalty is a reparameterization of the TRAC penalty \citep{bien2021tree}.

We also consider an $\ell_2$ penalty based on leaf-set containment, inspired by the descendant penalty of \citet{li2023s},
\[
\mathcal P_{\mathrm{des}}(\gamma)
=
\sum_{k=1}^T\| (\gamma_j)_{j\in D_A(k)}\|_2,\qquad D_A(k)=\{j\in\{1,\ldots,T\}:L_j\subseteq L_k\}.
\]
The sum covers all indexed nonroot nodes, including leaves, and each set contains its own node. This overlapping group-lasso penalty promotes joint sparsity within each containment set,
 and is included in the empirical comparisons.

In the numerical analyses, we center the response and observed log-ratio covariates using the fitting sample means. For a fitting sample of size $n$, we use centered cross-products and subtract $(n-1)\widehat\Sigma_U^{(p)}/n$ from the Gram matrix.
The leaf-space projection in \eqref{eq:leaf_projection_main} can be solved by an alternating direction method of multipliers
\citep{boyd2011distributed,datta2017cocolasso}. The gamma program is convex with linear constraints and can be solved using
standard convex optimization software such as \texttt{CVXR} \citep{CVXR}.

We select $\lambda$ by $K$-fold cross-validation. Both the training fit and the
held-out criterion use the kernel-preserving projection in
\eqref{eq:kernel_projection}, applied separately to the corrected Gram matrix
of the training folds and of the held-out fold. When the covariance is
estimated from auxiliary replicates, we hold this estimate fixed across the
folds of the primary data. We then fit TARCO to the full primary sample at the
selected $\hat{\lambda}$ and report
$\widehat\beta(\hat{\lambda})$. The fitted intercept on the original scale is $\bar y-\bar{\widetilde Z}^{\top}\widehat\beta$. Section~\suppref{sec:cv_appendix} details the training and validation calculations. For a specified threshold $\tau_\beta\geq0$, we form the maximal subtrees whose fitted coefficient range is at most $\tau_\beta$. Section~3.3 gives the formal definition and recovery guarantee.

In practice, we choose a relatively abundant reference component and replace observed zeros with a small pseudocount before applying the log-ratio transformation. Changing the reference applies an invertible linear transformation to
the additive log-ratio coordinates, and the error-free log-contrast criterion
is unchanged on $\cB$. The population invariance and the possible
finite-sample reference dependence of the projected criterion are summarized
in Remark~\ref{rem:reference_invariance}.

\Needspace{4\baselineskip}
\begin{remark}[Population invariance and finite-sample reference dependence]
\label{rem:reference_invariance}
For any $b\in\cB$, changing the additive log-ratio reference changes every
embedded row by a constant multiple of $\mathbf 1_p$. Hence the linear
predictor $Zb$ and the population quadratic form $b^\top Q_0b$, where
$Q_0=\mathbb E(n^{-1}Z^\top Z)$, are unchanged. The tree map and
$\Omega_\alpha$ are unchanged as well. Consequently, so are the beta-space
error set and compatibility constant introduced in
Section~\ref{sec:theoretical_results}.

The weighted positive semidefinite projection is a nonlinear finite-sample
operation and may depend on the chosen reference. Remark~\suppref{rem:pointwise_reference_comparison} bounds
the difference between the projected quadratic forms under two references
for a fixed $b\in\cB$.
This pointwise criterion bound does not imply that the two finite-sample
minimizer sets are identical.
\end{remark}

\section{Theoretical properties}
\label{sec:theoretical_results}

Under the conditions below on the design and penalty parameter, TARCO provides finite-sample bounds on prediction and coefficient estimation errors. Under the stated separation and threshold conditions, the bound on coefficient estimation error guarantees recovery of the true subtree groups after thresholding. We also extend the results to settings in which the measurement error covariance is estimated from auxiliary replicates. Proofs of the results in this section are given in Sections~\suppref{sec:common_event}--\suppref{sec:replicate_supp} of the supplementary materials.

\subsection{Stochastic assumptions and a common event}

Let $Q_Z=n^{-1}Z^\top Z$ and $Q_0=\mathbb E(Q_Z)$ denote the latent sample Gram matrix and its population counterpart, respectively.
Uniformly in $i$, assume
\begin{equation}
\max_{1\leq j\leq p}\|Z_{ij}\|_{\psi_2}\leq\nu_Z,
\qquad
\max_{1\leq j\leq p}\|U_{ij}\|_{\psi_2}\leq\nu_U.
\label{eq:tail_assumptions}
\end{equation}
The regression error satisfies
\begin{equation}
\mathbb E(e_i\mid Z_i)=0,
\qquad
\mathbb E\{\exp(te_i)\mid Z_i\}
\leq
\exp(t^2\sigma_e^2/2),
\qquad t\in\R.
\label{eq:error_tail}
\end{equation}
Here, $\nu_Z$ and $\nu_U$ describe the coordinatewise scales of the covariates and measurement errors, while $\sigma_e$ describes the regression error scale. The bounds in \eqref{eq:tail_assumptions} allow dependence among coordinates within each vector.

Set
\begin{align*}
\rho_{n,T}
&=
\sqrt{\frac{\log(eT)}{n}}
+\frac{\log(eT)}{n},\\
\zeta_1
&=
\max\{\nu_U^2\sigma_e^2,
\nu_U^2\nu_Z^2\|\beta^*\|_1^2\},
\qquad
\zeta_2=\max\{\nu_U^2\nu_Z^2,\nu_U^4\}.
\end{align*}
Also define
$L_{\max}(\alpha)=\max_k w_k^{1-\alpha}$ and $B_\alpha=\max_k w_k^{-\alpha}$.

For $\beta^*\neq0$, define the error set for $\beta$ by
\[
\cE_\alpha(\beta^*)
=
\left\{
h\in\cB:
\Omega_\alpha(\beta^*+h)-\Omega_\alpha(\beta^*)
\leq
\frac12\Omega_\alpha(h)
\right\}.
\]
We measure the deviation due to sampling by $\eta_Z$ and the additional deviation due to measurement error and its correction by $\eta_U$. Define
\begin{align*}
\eta_Z
&=
\sup_{h\in\cE_\alpha(\beta^*)\setminus\{0\}}
\frac{|h^\top(Q_Z-Q_0)h|}{\Omega_\alpha^2(h)},\\
\eta_U
&=
\sup_{h\in\cE_\alpha(\beta^*)\setminus\{0\}}
\frac{|h^\top(\widetilde Q-Q_Z)h|}{\Omega_\alpha^2(h)}.
\end{align*}

The next lemma controls the score residual $\widetilde q-\widetilde Q\beta^*$ and both quadratic deviations on a common event.
The proof uses a fixed representation $\gamma^\dagger$ satisfying $A\gamma^\dagger=\beta^*$, constructed in Section~\suppref{sec:fixed_comparison} of the Supplement. The bounds are stated in terms of $\beta^*$.

\begin{lemma}[Common stochastic event]
\label{lem:common_event}
Suppose $\beta^*\neq0$. Suppose also that the independence and moment
assumptions in
Section~\ref{sec:error_model}, together with
\eqref{eq:tail_assumptions} and \eqref{eq:error_tail}, hold and
$\widehat\Sigma_U^{(p)}=\Sigma_U^{(p)}$. There are positive constants
$C_0,C_1,c_0$, independent of $n,p$, and $T$, such that, with probability
at least $1-C_0T^{-c_0}$,
\begin{align}
\Omega_\alpha^*(\widetilde q-\widetilde Q\beta^*)
&\leq
C_1L_{\max}(\alpha)\rho_{n,T}
\left\{
\nu_Z\sigma_e+\sqrt{\zeta_1}
+\|\beta^*\|_1\sqrt{\zeta_2}
\right\},
\label{eq:score_bound}\\
\eta_Z
&\leq
C_1L_{\max}^2(\alpha)\nu_Z^2\rho_{n,T},
\qquad
\eta_U
\leq
C_1L_{\max}^2(\alpha)\sqrt{\zeta_2}\rho_{n,T}.
\notag
\end{align}
When $\nu_U=0$, the measurement error is zero almost surely,
$\eta_U=0$, and the score bound reduces to the term due to response noise
$C_1L_{\max}(\alpha)\nu_Z\sigma_e\rho_{n,T}$.
\end{lemma}

The bounds hold, with adjusted universal constants, for the sample-centered fit with covariance correction $(n-1)\Sigma_U^{(p)}/n$. The population matrix $Q_0$ is unchanged (Section~\suppref{sec:common_event}).

Under the same stochastic assumptions, inequality \eqref{eq:score_bound} also holds when $\beta^*=0$, with $\gamma^\dagger=0$ and $\|\beta^*\|_1=0$. The condition $\beta^*\neq0$ in Lemma~\ref{lem:common_event} is needed only for the definitions and bounds of $\eta_Z$ and $\eta_U$.

\subsection{Compatibility and oracle bounds}

To obtain estimation bounds, we also need the following condition linking the size of coefficient errors to the population quadratic form.

\begin{condition}[Beta-space compatibility]
\label{cond:beta_compatibility}
For $\beta^*\neq0$, define
\[
\kappa_\alpha^2(\beta^*;Q_0)
=
\inf_{h\in\cE_\alpha(\beta^*)\setminus\{0\}}
\frac{h^\top Q_0h}{\Omega_\alpha^2(h)}.
\]
Assume $\kappa_\alpha^2(\beta^*;Q_0)>0$.
\end{condition}

This condition is stated directly for $\beta$. Different tree representations of the same coefficient vector therefore do not affect it.

In the next theorem, $\lambda_{n,\alpha}$ is twice the score bound in Lemma~\ref{lem:common_event}, and $d_{n,\alpha}$ bounds $\eta_Z+\eta_U$. The condition on $d_{n,\alpha}$ preserves positive curvature on the error set. Define
\begin{align*}
\lambda_{n,\alpha}
&=
2C_1L_{\max}(\alpha)\rho_{n,T}
\left\{
\nu_Z\sigma_e+\sqrt{\zeta_1}
+\|\beta^*\|_1\sqrt{\zeta_2}
\right\},
\\
d_{n,\alpha}
&=
C_1L_{\max}^2(\alpha)
(\nu_Z^2+\sqrt{\zeta_2})\rho_{n,T}.
\end{align*}

\begin{theorem}[Prediction and estimation bounds]
\label{th:prediction_error}
Under the stochastic assumptions in Lemma~\ref{lem:common_event}, suppose
$\lambda>0$,
$\widehat\Sigma_U^{(p)}=\Sigma_U^{(p)}$, $\beta^*\neq0$,
Condition~\ref{cond:beta_compatibility} holds,
$\lambda\geq\lambda_{n,\alpha}$, and
$d_{n,\alpha}\leq\kappa_\alpha^2(\beta^*;Q_0)/2$.
Then, with probability at least $1-C_0T^{-c_0}$, the beta program
\eqref{eq:beta_estimator} has a minimizer and every minimizer satisfies
\[
\begin{aligned}
\Omega_\alpha(\widehat\beta-\beta^*)
&\leq
\frac{6\lambda}{\kappa_\alpha^2(\beta^*;Q_0)},\\
(\widehat\beta-\beta^*)^\top Q_0
(\widehat\beta-\beta^*)
&\leq
\frac{36\lambda^2}{\kappa_\alpha^2(\beta^*;Q_0)},
\qquad
\frac{\|Z(\widehat\beta-\beta^*)\|_2^2}{n}
\leq
\frac{36\lambda^2}{\kappa_\alpha^2(\beta^*;Q_0)},
\\
\|\widehat\beta-\beta^*\|_1
&\leq
\frac{6L_{\max}(\alpha)\lambda}
{\kappa_\alpha^2(\beta^*;Q_0)},
\qquad
\|\widehat\beta-\beta^*\|_\infty
\leq
\frac{6B_\alpha\lambda}
{\kappa_\alpha^2(\beta^*;Q_0)}.
\end{aligned}
\]
The same conclusions hold for every selected constrained projection and for
$A\widehat\gamma$ for every minimizer of the corresponding gamma program
\eqref{eq:TARCO}.
For the sample-centered fit, the same conclusions hold with the centered score and Gram matrices and with $n^{-1}\|H_nZ(\widehat\beta-\beta^*)\|_2^2$ as the sample prediction error, where $H_n=I_n-n^{-1}\mathbf 1_n\mathbf 1_n^\top$.
\end{theorem}

Theorem~\suppref{thm:deterministic_oracle_supp} in the Supplement gives the corresponding deterministic bounds in terms of $\eta_Z$ and $\eta_U$.

\begingroup\widowpenalty=10000
\begin{corollary}[Null target]
\label{cor:null_target_main}
Suppose the independence and moment assumptions in Section~\ref{sec:error_model} and conditions \eqref{eq:tail_assumptions} and \eqref{eq:error_tail} hold.
If $\beta^*=0$, $\lambda>0$, and $\lambda$ is at least twice the right-hand side of
\eqref{eq:score_bound}, evaluated at $\beta^*=0$, then, with probability
at least $1-C_0T^{-c_0}$, every beta minimizer and the beta image of every
gamma minimizer equal zero.
\end{corollary}
\endgroup

\subsection{Beta-level group recovery}

The bound on the largest coefficient error in Theorem~\ref{th:prediction_error} also controls errors in the coefficient range within each subtree. We use this bound to show when thresholding these ranges recovers the true groups.

Define the collection of tree nodes that merge at least two true groups by
\[
\cM
=
\left\{
u\in\mathcal V(\mathcal T):
L_u\text{ contains at least two distinct blocks of }\cG^*
\right\},
\]
and let $\Delta_{\mathrm{merge}}$ be the smallest range of true coefficients among these nodes:
\[
\Delta_{\mathrm{merge}}
=
\begin{cases}
\displaystyle
\min_{u\in\cM}\diam_{\beta^*}(L_u),
&\cM\neq\varnothing,\\[6pt]
+\infty,
&\cM=\varnothing.
\end{cases}
\]
For a threshold $\tau_\beta\geq0$, define the estimated groups as the maximal subtrees whose estimated coefficient range is at most $\tau_\beta$:
\[
\widehat\cG(\tau_\beta)
=
\left\{
L_k:
\begin{array}{l}
k\in\mathcal V(\mathcal T),\\
\diam_{\widehat\beta}(L_k)\leq\tau_\beta,\\
\text{there is no strict ancestor }a\succ k
\text{ with }\diam_{\widehat\beta}(L_a)\leq\tau_\beta
\end{array}
\right\}.
\]

\begin{theorem}[Tree-constrained group recovery for $\beta^*$]
\label{th:tree_group_recovery}
Under the assumptions of Theorem~\ref{th:prediction_error}, set
$r_n=6B_\alpha\lambda/\kappa_\alpha^2(\beta^*;Q_0)$.
If
\[
\Delta_{\mathrm{merge}}>4r_n,
\qquad
2r_n\leq\tau_\beta<\Delta_{\mathrm{merge}}-2r_n,
\]
then, with probability at least $1-C_0T^{-c_0}$,
$\widehat\cG(\tau_\beta)=\cG^*$ for every beta minimizer and for the beta
image of every gamma minimizer.
\end{theorem}

The stated threshold range ensures that each true group passes the diameter threshold and every subtree merging distinct groups fails it. Thresholding $\widehat\beta$ therefore recovers the groups defined by $\beta^*$, even though the estimated coefficients within a group need not be exactly equal.

\subsection[Extension to estimated measurement error covariance]{Extension to estimated measurement error covariance}
\label{sec:replicate}

When $\Sigma_U^{(p)}$ is unknown, we use the unbiased estimator below based on auxiliary replicates. The primary and auxiliary samples need not be independent. Suppose an auxiliary collection contains $n_a$
independent replicate units,
with fixed counts $t_i\geq2$ in unit $i$. Write
$\widetilde Z_{it}^{\mathrm{aux}}=Z_i^{\mathrm{aux}}+U_{it}^{\mathrm{aux}}$,
$t=1,\ldots,t_i$,
where $Z_i^{\mathrm{aux}}$ is shared by all replicates from unit $i$. Assume the replicate
errors are independent within each unit, have mean zero and common covariance
$\Sigma_U^{(p)}$, are independent across units, and satisfy the uniform
sub-Gaussian bound in \eqref{eq:tail_assumptions} with scale $\nu_U$.
Let $\overline Z_{i\cdot}^{\mathrm{aux}}
=t_i^{-1}\sum_{t=1}^{t_i}\widetilde Z_{it}^{\mathrm{aux}}$ and define
\begin{equation}
\widehat\Sigma_U^{(p)}
=
\frac{
\sum_{i=1}^{n_a}\sum_{t=1}^{t_i}
(\widetilde Z_{it}^{\mathrm{aux}}-\overline Z_{i\cdot}^{\mathrm{aux}})
(\widetilde Z_{it}^{\mathrm{aux}}-\overline Z_{i\cdot}^{\mathrm{aux}})^\top
}
{\sum_{i=1}^{n_a}(t_i-1)}.
\label{eq:covariance_estimator}
\end{equation}

\begin{corollary}[Estimated covariance]
\label{cor:estimated_covariance_matrix}
Assume that the primary data satisfy the measurement error model and the independence and moment assumptions in Section~\ref{sec:error_model}, together with \eqref{eq:tail_assumptions} and \eqref{eq:error_tail}. Suppose $\beta^*\neq0$ and
Condition~\ref{cond:beta_compatibility} holds. Also impose the
assumptions on auxiliary replicates above and suppose there is a constant $C_r$
such that
\mbox{$\max_i(t_i-1)\leq C_r\sum_{i=1}^{n_a}(t_i-1)/n_a$}.
Let $\rho_{n_a,T}=\sqrt{\log(eT)/n_a}+\log(eT)/n_a$.
There are positive constants $C_{\lambda,\mathrm{rep}}$,
$C_{d,\mathrm{rep}}$, $C_{\mathrm{rep},0}$, and $c_{\mathrm{rep}}$, depending
at most on $C_r$ and universal constants, such that the following holds.
Estimating the covariance adds the following terms to $\lambda_{n,\alpha}$
and $d_{n,\alpha}$, respectively:
\[
C_{\lambda,\mathrm{rep}}L_{\max}(\alpha)\|\beta^*\|_1
\nu_U^2\rho_{n_a,T},
\qquad
C_{d,\mathrm{rep}}L_{\max}^2(\alpha)\nu_U^2\rho_{n_a,T}.
\]
If $\lambda>0$, $\lambda$ is at least $\lambda_{n,\alpha}$ plus the first
term, and $d_{n,\alpha}$ plus the second term is at most
$\kappa_\alpha^2(\beta^*;Q_0)/2$, then, with probability at least
$1-C_{\mathrm{rep},0}T^{-c_{\mathrm{rep}}}$, the prediction and estimation
bounds in Theorem~\ref{th:prediction_error} hold for every selected
replicate-corrected projection and every corresponding minimizer. Under the
separation and threshold conditions of Theorem~\ref{th:tree_group_recovery}, its
group recovery conclusion holds on the same event.
For the sample-centered fit with the replicate covariance, the sample prediction error is measured as in the centered version of Theorem~\ref{th:prediction_error}.
\end{corollary}

Suppose the auxiliary sample size $n_a$ grows at the same rate as $n$, the replicate counts satisfy the stated balance condition, and both samples have the same error scale $\nu_U$. The additional covariance estimation terms are then of no larger order than $\lambda_{n,\alpha}$ and $d_{n,\alpha}$, respectively.
 
\section{Simulation studies} \label{sec:simulation_studies}

The simulations evaluate how measurement error correction affects prediction, coefficient estimation, and group recovery in tree-aggregated regression. We also examine how penalty weights across tree levels affect performance. We compare TARCO with TRAC \citep{bien2021tree} and our ALR implementation of CoCoLasso, denoted by COCO \citep{datta2017cocolasso}. TRAC uses tree aggregation without correcting measurement error, whereas COCO corrects measurement error without tree aggregation. TRAC uses unit node weights, giving the penalty $\sum_{k=1}^T|\gamma_k|$ ($\alpha=0$). Thus, TRAC and the main TARCO fit use the same tree penalty. We also compare penalties based on leaf counts with $\alpha=0$ (TARCO), $\alpha=0.5$ (TARCO-05), and $\alpha=-0.5$ (TARCO-n05), together with the $\ell_2$ penalty variant (TARCO-Des).

\subsection{Simulation settings}

We consider a baseline setting with $(n,p)=(100,100)$, a setting with more variables, $(n,p)=(100,200)$, and a setting with a larger sample, $(n,p)=(500,100)$. In each setting, we generate compositional data $X = (X_{ij}) \in \mathbb{R}^{n \times p}$ from a logistic-normal distribution following \citet{lin2014variable}. Specifically, we first draw $G = (g_{ij}) \in \mathbb{R}^{n \times p}$ from a multivariate normal distribution with mean $\mu = (\mu_1, \ldots, \mu_p)^\top$ and covariance $\Sigma_{ij} = \rho^{|i-j|}$, where $\rho = 0.5$. Following \citet{li2023s}, we set $\mu_j = \log(p/2)$ for $j = 1, \ldots, 5$ and $\mu_j = 1$ otherwise. The compositional data are then obtained by the softmax transformation $X_{ij} = \exp(g_{ij}) / \sum_{l=1}^p \exp(g_{il})$.

The log-ratio transformation yields $Z = \{\log(X_{ij}/X_{ip})\} \in \mathbb{R}^{n \times (p-1)}$, with the $p$th component as the reference. Here we describe the $p-1$ non-reference coordinates; the zero reference column is added as in Section~\ref{sec:model_setup}. The rows of the measurement error matrix $U$ are drawn independently from a multivariate normal distribution with mean zero and covariance $\Sigma_U = \tau^2(\mathbf I_{p-1} + \mathbf 1_{p-1}\mathbf 1_{p-1}^\top)$, with $\tau = 1$. Section~\suppref{sec:toy_example_appendix} explains this covariance structure. The corrupted data are $\widetilde{Z} = Z + U$.

The true coefficient vector is
$\beta^* = (0.5 \mathbf{1}_{p/5}^\top,\allowbreak {-}0.75 \mathbf{1}_{p/10}^\top,\allowbreak
{-}0.25 \mathbf{1}_{p/10}^\top,\allowbreak 0.1 \mathbf{1}_{p/5}^\top,\allowbreak
{-}0.1 \mathbf{1}_{p/5}^\top,\allowbreak \mathbf{0}_{3p/20}^\top,\allowbreak
\nu_{p/20}^\top)^\top$,
where $\mathbf{1}_q$ denotes a vector of $q$ ones and $\nu_{p/20}$ is drawn from $N(0,0.25I_{p/20})$ and then centered. The centered vector is generated once for each value of $p$ and held fixed across Monte Carlo replicates. The first five constant blocks define effects shared within subtrees, including two weak blocks with coefficients $0.1$ and $-0.1$, while the remaining coordinates are represented at the leaf level. The vector satisfies $\mathbf{1}_p^\top\beta^*=0$. In the three principal settings, 85\% of the leaf coefficients are nonzero. For $p=100$, the first five constant blocks are maximal constant subtrees under Definition~\ref{def:aggre_process}, while the remaining 20 leaves form singleton groups, giving $|\mathcal G^*|=5+20=25$ (Figure~\ref{fig:TreeSimulation}).   Responses are generated from the linear model~\eqref{eq:logratio_linear}, with independent errors $e_i\sim N(0,0.5^2)$.

We also consider a setting with only singleton groups. With $(n,p)=(100,100)$, we break the constant coefficient blocks by setting $\beta_j^*=0$ for $j=1,6,11,\ldots,76$.

For the results in Table~\ref{tab:sim_results}, we estimate the measurement error covariance from $n$ auxiliary units, each with two replicates. The two replicates share the same latent log-ratio vector and have independent measurement errors. These $2n$ observations are used only to compute $\widehat{\Sigma}_U$ through~\eqref{eq:covariance_estimator}. TRAC uses the observed compositions obtained by applying the inverse ALR transformation to the primary log-ratio data. COCO is fit to $(\widetilde Z,y)$ using the same $\widehat\Sigma_U$, with its $(p-1)$-dimensional estimate expanded to satisfy the log-contrast constraint. All tuning parameters are selected by 5-fold cross-validation. All methods use training-sample centering to account for an unpenalized intercept.

To assess the contribution of tree aggregation, we construct a comparator without tree aggregation, denoted by MatchedNoTree. It uses TARCO's corrected loss and PSD projection in ALR coordinates, but applies an $\ell_1$ penalty directly to the leaf coefficients. We also evaluate Eric Lasso, which performs the correction and PSD projection in the full log-composition space (\citeauthor{tan2024high}, \citeyear{tan2024high}). We use known measurement error covariance for TARCO and both comparators to exclude covariance estimation error from this comparison.

\begingroup\widowpenalty=10000
We assess performance using three metrics. To evaluate prediction on the original response scale, we set $\hat b_0=\bar y-\bar{\widetilde Z}^{\top}\hat\beta$, where the means are computed from the full training sample. Predictive accuracy is measured by the mean squared prediction error
$\mathrm{MSPE}(\hat\beta)
=\frac1n\left\|y_{\mathrm{test}}-\hat b_0\mathbf1_n-\log(X_{\mathrm{test}})\hat\beta\right\|_2^2$
on an independent test sample of size $n$, using the latent, error-free compositions. Absolute error, $\mathrm{AE}(\hat{\beta}) = \|\hat{\beta} - \beta^*\|_1$, quantifies estimation accuracy. Group recovery, denoted by $\mathrm{GR}(\hat\beta)$, is the adjusted Rand index (ARI) \citep{hubert1985comparing} between the true partition $\mathcal G^*$ and the estimated partition $\widehat{\mathcal G}(\tau_\beta)$. Lower MSPE and AE indicate better performance, whereas larger GR is better and GR equal to 1 indicates exact agreement. We set $\tau_\beta=0.05$ in the diameter rule of Theorem~\ref{th:tree_group_recovery} and apply the same grouping rule constrained by the tree to all methods. \mbox{Results are based on 100 Monte Carlo replicates.}\par\endgroup

\begin{figure}[htbp]
\centering

\begin{subfigure}[b]{0.48\linewidth}
  \centering
  \includegraphics[width=\linewidth,height=5cm,keepaspectratio]{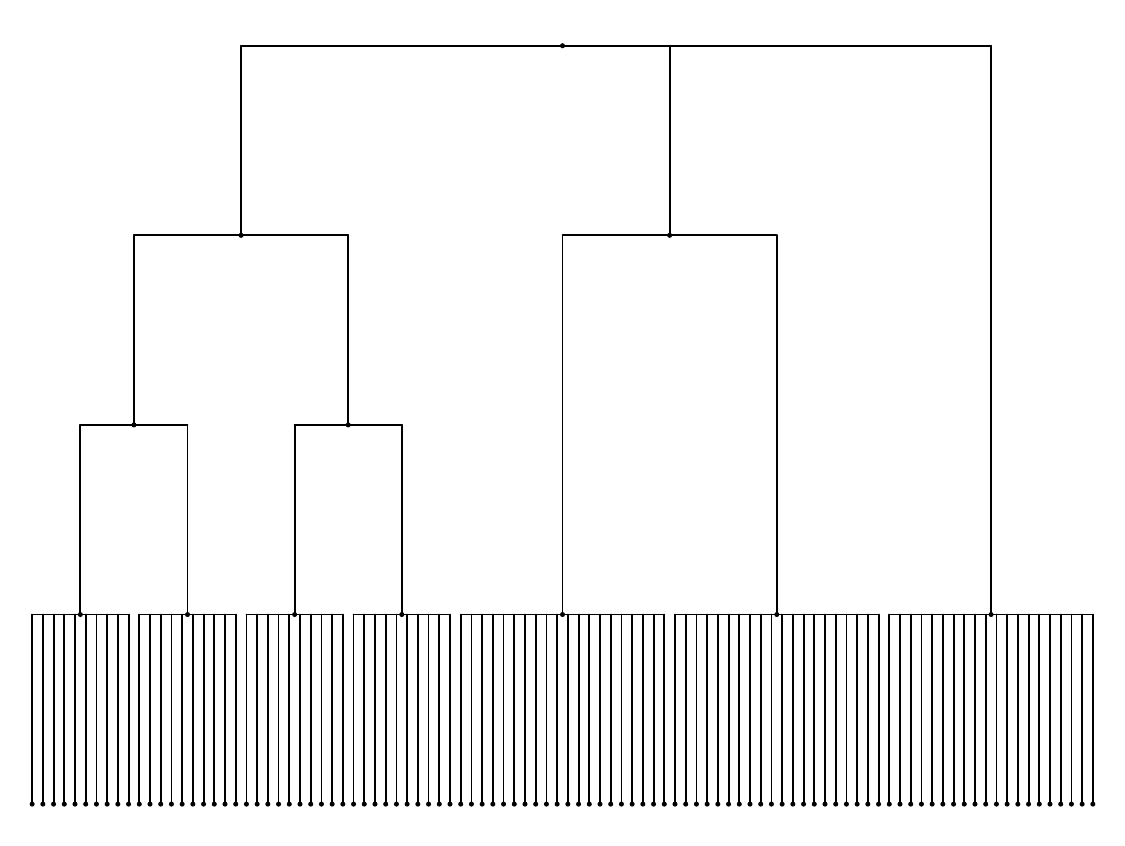}
  \caption{}
\end{subfigure}\hfill
\begin{subfigure}[b]{0.48\linewidth}
  \centering
  \includegraphics[width=\linewidth,height=5cm,keepaspectratio]{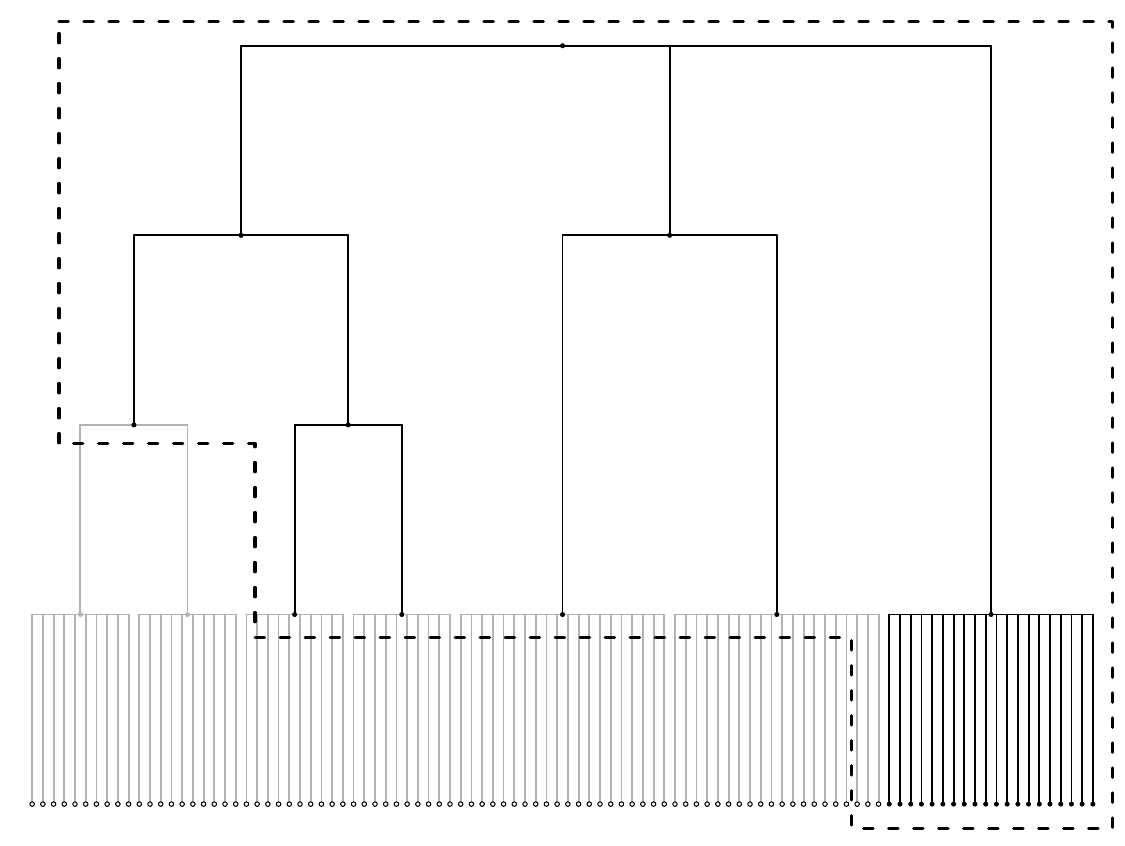}
  \caption{}
\end{subfigure}

\renewcommand{\baselinestretch}{1}\selectfont
\caption{Simulation tree before (a) and after (b) aggregation, with $p=100$ leaves. Gray branches indicate the collapsed parts, and dashed outlines mark the aggregation levels.}
\label{fig:TreeSimulation}
\end{figure}
\FloatBarrier

\subsection{Simulation results}

Table~\ref{tab:sim_results} reports results with estimated measurement error covariance. With the tree penalty held fixed, TARCO has lower mean MSPE and AE than TRAC in all three principal settings. Its mean GR is higher in the two settings with $n=100$ and comparable to TRAC's when $n=500$. All four TARCO variants also have lower mean MSPE and AE and higher mean GR than COCO in these settings. For TARCO with $p=100$, GR is lower at $n=500$ despite the smaller AE. At the fixed grouping threshold, more pairs from the same true group are assigned to different estimated groups. Table~\suppref{tab:benchmark_threshold} reports sensitivity to this threshold.
When all true groups are singletons, all four TARCO variants have lower mean MSPE and AE than TRAC and COCO. Here, GR equals 1 for exact partition recovery and 0 otherwise, so its mean is the proportion of replicates with exact recovery. Section~\suppref{sec:additional_simulation_appendix} of the Supplement examines sensitivity to the measurement error parameter $\tau$ and the grouping threshold $\tau_\beta$.

Among the penalty variants, TARCO has the lowest mean AE in all three principal settings and the lowest mean MSPE in the two settings with $n=100$. TARCO-n05 places less penalty on larger nodes and has the highest mean GR throughout, with slightly higher mean MSPE and AE than TARCO. TARCO-Des gives a slightly lower mean MSPE than TARCO when $(n,p)=(500,100)$, but its mean GR is lower. Thus, the penalty giving the best prediction need not give the most accurate group recovery.

When the error covariance is known, TARCO has lower mean MSPE and AE and higher mean GR than MatchedNoTree and Eric Lasso in all three principal settings. The comparison with MatchedNoTree demonstrates the additional benefit of tree aggregation when the corrected loss is held fixed. Results for the additional comparators are reported in Tables~\suppref{tab:extra_comparators_main} and~\suppref{tab:extra_comparators_tau} of the Supplement.

\begin{table}[!htbp]
\centering
\renewcommand{\baselinestretch}{1}\selectfont
\caption{Simulation results with estimated measurement error covariance. Entries are Monte Carlo mean (SD) over 100 replicates.}
\label{tab:sim_results}
\renewcommand{\baselinestretch}{1.5}
\small
\setlength{\tabcolsep}{2.5pt}
\renewcommand{\arraystretch}{0.88}
\resizebox{\textwidth}{!}{
\begingroup
\begin{tabular}{@{}lcccccc@{}}
\toprule
Metric & TARCO & TARCO-05 & TARCO-n05 & TARCO-Des & TRAC & COCO \\
\midrule
\multicolumn{7}{c}{\textbf{$p=100$, $n=100$}} \\
MSPE & 2.48 (0.94) & 3.40 (1.23) & 2.63 (1.17) & 2.66 (1.02) & 4.55 (1.31) & 22.03 (6.73) \\
AE & 7.19 (1.80) & 9.45 (1.74) & 7.34 (2.08) & 7.63 (1.88) & 10.25 (1.47) & 24.66 (1.53) \\
GR & 0.76 (0.12) & 0.59 (0.17) & 0.78 (0.10) & 0.72 (0.16) & 0.68 (0.25) & 0.34 (0.22) \\
\midrule
\multicolumn{7}{c}{\textbf{$p=200$, $n=100$}} \\
MSPE & 5.14 (2.80) & 6.25 (2.69) & 5.72 (3.26) & 5.80 (2.65) & 7.83 (2.42) & 59.70 (11.74) \\
AE & 15.13 (4.05) & 18.06 (3.49) & 15.52 (4.37) & 16.28 (3.97) & 18.67 (2.85) & 51.09 (1.68) \\
GR & 0.74 (0.11) & 0.59 (0.14) & 0.78 (0.10) & 0.71 (0.13) & 0.70 (0.24) & 0.32 (0.24) \\
\midrule
\multicolumn{7}{c}{\textbf{$p=100$, $n=500$}} \\
MSPE & 1.23 (0.32) & 1.50 (0.38) & 1.26 (0.34) & 1.20 (0.31) & 3.45 (0.51) & 7.44 (1.38) \\
AE & 4.71 (1.12) & 6.12 (1.22) & 4.75 (1.15) & 5.02 (1.07) & 8.90 (0.70) & 20.50 (2.00) \\
GR & 0.66 (0.24) & 0.41 (0.26) & 0.70 (0.23) & 0.44 (0.27) & 0.66 (0.31) & 0.07 (0.15) \\
\midrule
\multicolumn{7}{c}{\textbf{Singleton groups ($p=100$, $n=100$)}} \\
MSPE & 3.32 (0.77) & 3.97 (1.09) & 3.50 (0.95) & 3.42 (0.82) & 4.63 (1.07) & 15.56 (4.54) \\
AE & 11.25 (1.33) & 12.94 (1.08) & 11.24 (1.53) & 11.48 (1.35) & 13.29 (0.99) & 19.94 (1.39) \\
GR & 0.00 (0.00) & 0.00 (0.00) & 0.00 (0.00) & 0.00 (0.00) & 0.05 (0.22) & 0.12 (0.33) \\
\bottomrule
\end{tabular}\endgroup
}
\end{table}
\FloatBarrier

\section{Application} \label{sec:applications}

We use longitudinal gut microbiome data to study associations between microbial composition and body mass index (BMI). Participants provided weekly samples over three months in a study of temporal variability in the human microbiome \citep{flores2014temporal}. Following the preprocessing and sample selection of \citet{shi2022high}, we analyze 160 samples from 40 participants and 89 genera. We ask whether associations with BMI are specific to individual genera or shared across broader taxonomic groups.

We compare TARCO with TRAC and COCO to examine how measurement error correction and tree aggregation affect these associations. TRAC uses the observed compositions without measurement error correction, whereas COCO corrects measurement error without tree aggregation. We use default TARCO, with $\alpha=0$, to hold the tree penalty fixed in the primary comparison with TRAC. It also has the lowest mean estimation error in the three principal simulation settings. We consider TARCO-05, TARCO-n05, and TARCO-Des to assess sensitivity to penalty choice.

The preprocessing excludes genera present in at most 10\% of samples. Following the Bray-Curtis screening of \citet{shi2022high}, four similar samples are selected for each participant. We treat these as repeated observations of an approximately common underlying composition. Zero counts are replaced by 0.1 before normalization, and the most abundant genus, \textit{Bacteroides}, serves as the ALR reference. All four observations per participant contribute to the working measurement error covariance estimate in \eqref{eq:covariance_estimator}.

We fit each method four times. Each fit uses one of the four selected samples from every participant, giving 40 observations. BMI and ALR covariates are centered within each training sample to accommodate an intercept. Each fit's penalty strength is selected by pooling validation losses across 20 repetitions of five-fold cross-validation. We report mean coefficients across the four fits and obtain groups by applying the diameter threshold of $0.01$ to these means.

Measurement error correction changes the relative strengths of the estimated BMI associations (Figure~\ref{realdata_combined}). Default TARCO and TRAC both give negative coefficients to 13 genera labeled \textit{Gammaproteobacteria} and positive coefficients to 21 genera labeled \textit{Clostridia}. Both methods show weaker associations for \textit{Bacilli} and \textit{Bacteroidia}. Within \textit{Gammaproteobacteria}, the eight \textit{Enterobacterales} genera share one coefficient and the remaining five genera share another under both methods. Relative to TRAC, TARCO gives \textit{Enterobacterales} a more negative coefficient and the other five genera a coefficient closer to zero. It also gives the \textit{Clostridia} group a larger positive coefficient. At the diameter threshold of $0.01$, TARCO gives 18 subtree groups, compared with 15 for TRAC. COCO gives genus-level estimates, with the largest absolute coefficients for \textit{Bacteroides} and \textit{Prevotella}. In this ALR implementation, the \textit{Bacteroides} coefficient is determined by the zero-sum constraint and is not penalized.

The positive association shared by the 21 \textit{Clostridia} genera also appears under all three alternative TARCO penalties, with mean coefficients close to $0.010$ under all four choices. Penalty choice affects the additional associations and their taxonomic resolution. TARCO-n05 makes larger nodes less costly. It gives the \textit{Bacilli} group a more negative coefficient than default TARCO, while its \textit{Gammaproteobacteria} coefficients are close to zero. TARCO-05 penalizes larger nodes more heavily and identifies associations at a finer taxonomic level within \textit{Enterobacteriaceae}. It also identifies genus-level associations for \textit{Lactobacillus} and \textit{Prevotella}. TARCO-Des gives negative coefficients to both \textit{Bacilli} and \textit{Gammaproteobacteria}. Its \textit{Gammaproteobacteria} coefficients are more similar to one another than those of default TARCO. Table~\suppref{tab:application_groups_v1} reports the coefficients, and Section~\suppref{sec:additional_application_appendix} describes the highlighting criteria and examines sensitivity to the ALR reference.

The positive \textit{Clostridia} group includes \textit{Blautia}, \textit{Coprococcus}, \textit{Eubacterium}, and \textit{Ruminococcus}. Species within these genera have been reported to be more abundant in individuals with obesity \citep{kasai2015comparison}. A positive association between the \textit{Blautia}/\textit{Bacteroides} ratio and BMI has also been reported \citep{kim2022association}. Across all four penalties, TARCO identifies these four genera as part of a broader positive BMI association.

\begin{figure}[!htbp]
    \centering
    \includegraphics[width=\textwidth]{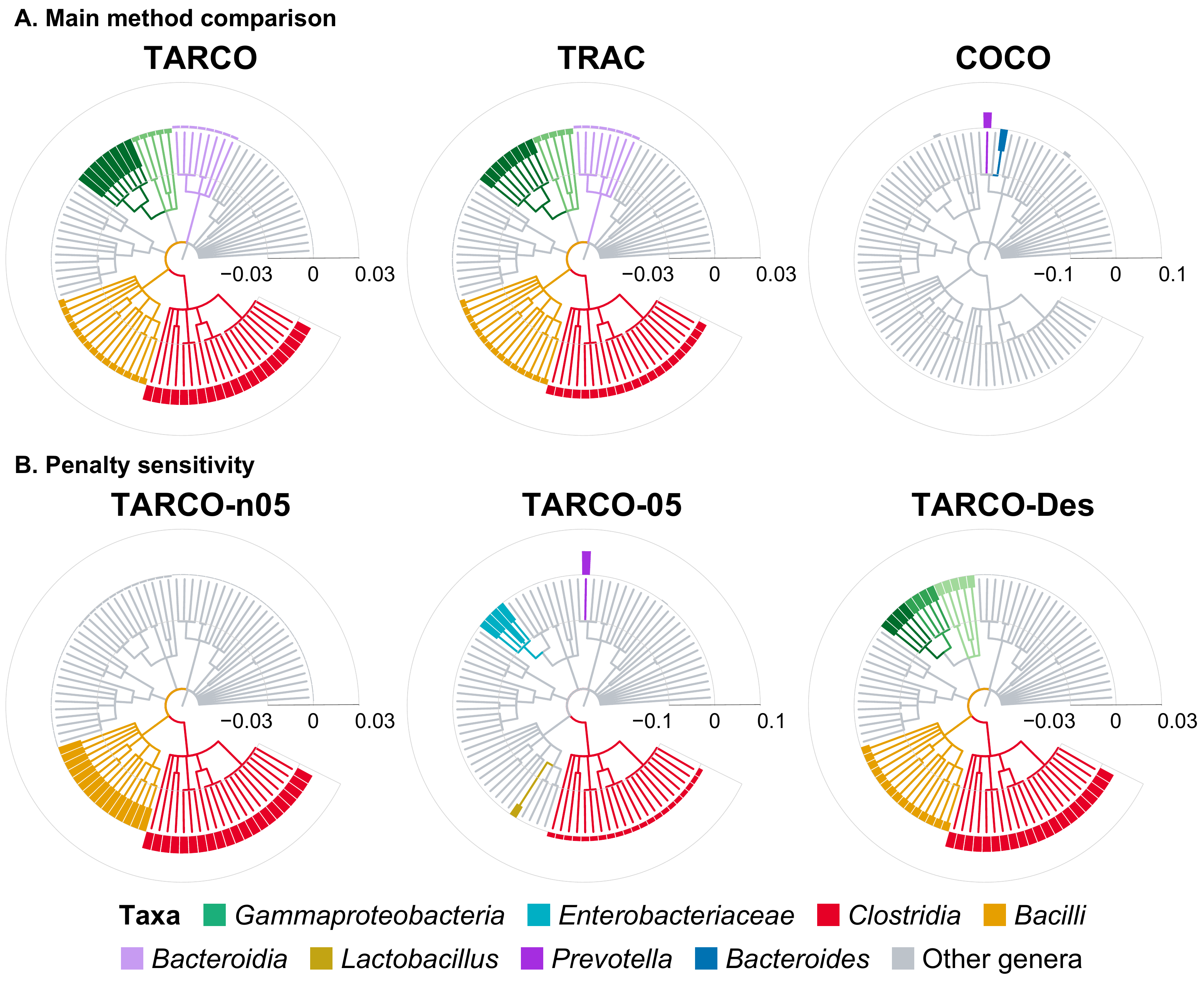}
    \renewcommand{\baselinestretch}{1.85}\selectfont
    \caption{Mean coefficients across four measurement-specific fits, using \textit{Bacteroides} as the ALR reference. Within \textit{Gammaproteobacteria}, different green shades indicate different mean coefficients for each method.}
    \label{realdata_combined}
\end{figure}
\FloatBarrier

\section{Discussion}

TARCO estimates log-contrast coefficients from error-prone compositions and uses them to identify subtrees with a shared coefficient. Our finite-sample theory establishes when estimation accuracy is sufficient to recover the latent subtree partition. This extends the analysis from individual coefficients to the level at which an association is shared. In the microbiome application, TRAC and all four TARCO penalties identify a shared positive BMI association within \textit{Clostridia}. Measurement error correction changes the relative strengths of the estimated associations, with the negative association within \textit{Gammaproteobacteria} more concentrated in \textit{Enterobacterales} under default TARCO.

\begingroup\widowpenalty=10000
Future work could examine how to select the tree penalty when group recovery is the primary objective. Our simulations show that the penalty giving the best prediction need not give the most accurate group recovery. The stability of selected groups under resampling may help guide this choice. Another direction is statistical inference for regression effects. Debiasing ideas developed for error-prone log-contrast regression could be adapted to TARCO to construct confidence intervals for individual coefficients and prespecified contrasts \citep{zhao2024debiased}. \mbox{Such inference} would complement partition recovery by quantifying uncertainty in the \mbox{estimated associations.}\par\endgroup

\subsection*{Data availability statement}

The processed data analyzed in this study are publicly available in the supplementary materials of \cite{shi2022high}.
\subsection*{Supplementary Material}

Supplementary materials provide proofs, additional numerical results, and implementation details. R code for TARCO is available at \url{https://github.com/tianyingw/TARCO}.

\bibliographystyle{agsm_doi}
\bibliography{paper_refer}

\endgroup
\end{document}